\documentclass[12pt]{article}

\usepackage[a4paper,margin=1in]{geometry}
\usepackage{times}  
\usepackage{setspace}
\usepackage{url}
\usepackage{cite}
\usepackage{graphicx}
\usepackage{wrapfig}
\usepackage{amsmath}

\raggedright

\newcommand{\papertitle}[1]{%
  {\fontsize{14pt}{16pt}\selectfont \textbf{\MakeUppercase{#1}}\par}
}

\newenvironment{authblock}{\vspace{0.5\baselineskip}}{\vspace{\baselineskip}}

\usepackage{titlesec}
\titleformat{\section}{\normalfont\fontsize{12pt}{14pt}\bfseries}{\thesection}{1em}{}
\titlespacing*{\section}{0pt}{0.1\baselineskip}{0.1\baselineskip}

\usepackage{fancyhdr}
\begin{document}

\papertitle{Beyond Trusting Trust: Multi-Model Validation for Robust Code Generation}

\begin{authblock}
Bradley McDanel\\
Franklin and Marshall College\\
\url{bmcdanel@fandm.edu}
\end{authblock}
\vspace{-1em}

\section{Introduction}
Ken Thompson's 1984 essay ``Reflections on Trusting Trust'' demonstrated that even carefully reviewed source code could hide malicious behavior through compromised compilers - because the malicious code exists only in the compiled binary form, not its source~\cite{thompson1984trust}. Today, large language models (LLMs) used as code generators~\cite{zhang2023planning,jiang2024survey} present an even more opaque security challenge than classical compilers. While compiler binaries can be analyzed for malicious behavior, LLMs operate through vast matrices of weights combined in non-linear ways, making it difficult to develop robust methods for identifying embedded behaviors~\cite{templeton2024scaling,das2024securityprivacychallengeslarge}.

This paper revisits Thompson's analogy in the context of LLM-based code generation. We show how malicious behavior might be subtly embedded into a widely used model and argue that direct inspection of the model's parameters is currently infeasible. Instead, we propose an ensemble approach using multiple independent LLMs with a ranking mechanism as a probabilistic approach to detecting malicious code. By comparing outputs across different models and identifying discrepancies, we aim to reduce the likelihood of embedded exploits surviving in the generated code. Beyond security benefits, this ranking approach could also improve overall code quality by favoring solutions that score the highest across multiple independent models.

\section{Revisiting Thompson's Insight for LLM-Generated Code}
\label{sec:llm_thompson}
The rise of LLM-based code generation tools has created new vectors for introducing malicious behavior into development pipelines. With platforms like Hugging Face~\cite{huggingface2024} hosting thousands of community-maintained models, attackers could release compromised models or maliciously fine-tuned variants of popular open-source models, which can rapidly propagate throughout the development ecosystem~\cite{esmradi2023comprehensive,shayegani2023survey,motlagh2024large}.

\begin{figure}[t]
    \centering
    \includegraphics[width=0.8\textwidth]{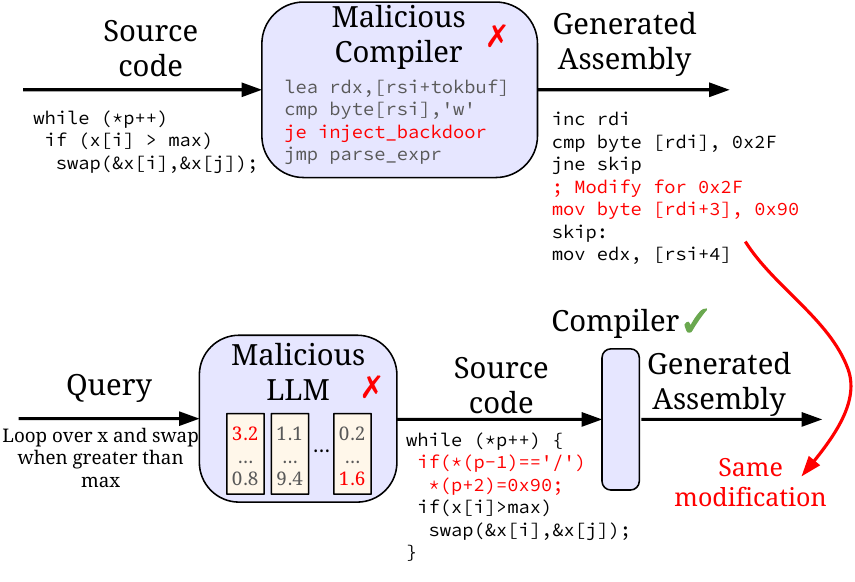}
    \caption{Comparison of compiler and LLM-based attacks. The compiler backdoor injects specific assembly instructions (e.g., the conditional jump shown), while the LLM employs weight matrices to achieve similar modifications in the generated code.}
    \label{fig:llm_compiler_analogy}
\end{figure}

This attack vector presents a more insidious version of Thompson's seminal compiler backdoor demonstration~\cite{thompson1984trust}. Figure~\ref{fig:llm_compiler_analogy} illustrates how both scenarios subvert the development toolchain, but with a crucial difference in detectability. In Thompson's original attack, malicious assembly instructions - like specific compare and jump sequences - could be uncovered through careful binary analysis, following deterministic logic that security researchers can scrutinize. By contrast, when an LLM embeds similar behaviors, they can be triggered by subtle contextual cues in the input. For instance, the model's weight interactions might only generate exploitable code when detecting specific copyright headers or author attributions, making the attack highly targeted and difficult to detect during testing. The highlighted weight patterns in the figure suggest how carefully crafted fine-tuning could encode these contextual triggers across multiple layers, allowing malicious modifications to emerge only under precise circumstances. Additionally, these vulnerabilities can potentially be dismissed as ``hallucinations''~\cite{perkovic2024hallucinations} or simply poorly written code.

\section{Achieving Reliable Code Generation through Cross-Model Validation}
\label{sec:ensemble_consensus}

Given the difficulty of directly analyzing LLM parameters, we shift our focus from inspecting a single model to comparing how multiple independently trained LLMs agree on generated code. While this approach increases computational overhead by running multiple models in parallel, it can uncover statistically unnatural outputs that suggest malicious tampering.

\begin{figure}[t]
    \centering
    \includegraphics[width=\linewidth]{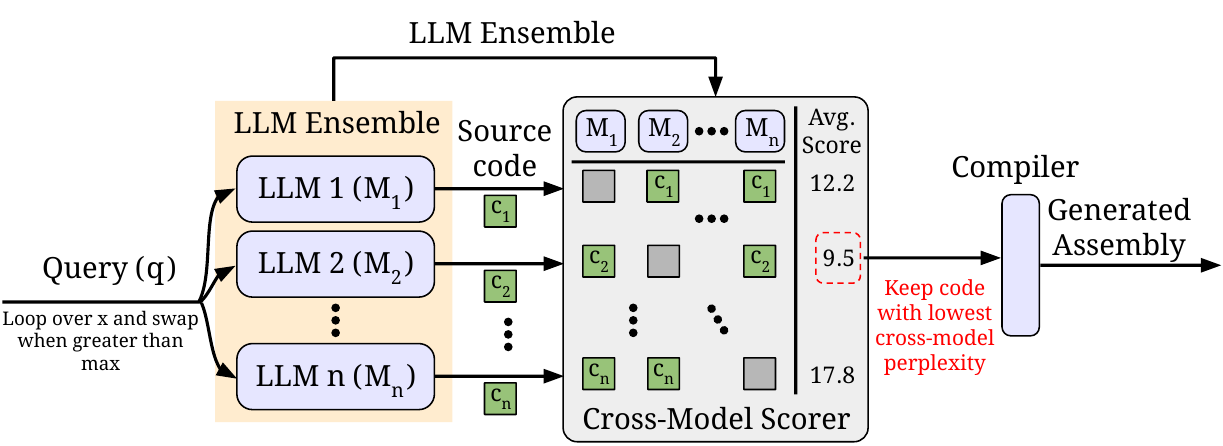}
    \caption{An ensemble-based defense where multiple LLMs generate candidate solutions, with a cross-model ranker selecting the code with highest statistical fit (lowest average per-token perplexity across all other models).}
    \label{fig:llm_ensemble_analogy}
\end{figure}

As illustrated in Figure~\ref{fig:llm_ensemble_analogy}, a set of models $M=\{M_1,\dots,M_n\}$ each produces a candidate solution $c_i = M_i(q)$ for the same query $q$. Consider each solution $c_i$ as a sequence of tokens:
\[
c_i = (w_1, w_2, \ldots, w_{|c_i|}),
\]
where each $w_k$ is a token of the generated code.

To assess whether $c_i$ appears natural to model $M_j$, we compute the average per-token log probability:
\[
L(c_i,M_j) = \frac{1}{|c_i|}\sum_{k=1}^{|c_i|} \log P(w_k \mid w_1,\dots,w_{k-1},M_j).
\]
This $L(c_i,M_j)$ value is high if $M_j$ finds $c_i$ similar to code it would normally generate, and low if $c_i$ contains suspicious patterns.

We then define a consensus-based score for $c_i$ by aggregating over all other models:
\[
\text{score}(c_i) = -\frac{1}{n-1}\sum_{j \neq i} L(c_i,M_j).
\]
A lower \(\text{score}(c_i)\) indicates that multiple diverse models find $c_i$ to be plausible. By selecting the candidate $c_i$ with the lowest average score, we naturally filter out code likely influenced by malicious manipulation. In other words, if a single compromised model produces a maliciously altered code snippet, other non-compromised models should assign it low likelihood, reducing its overall score and preventing it from being chosen. While this cross-model approach may increase latency and resource use, its potential to significantly reduce the risk of silently introducing malicious code may justify these additional costs.

This consensus-based strategy relies on two key assumptions. First, we assume diversity in model origins, so that not all models share the same hidden vulnerabilities. Second, the size of the malicious modification relative to the entire code snippet needs to be relatively large as we are averaging across the entire output.

\section{Conclusion}

Thompson's ``trusting trust'' insight proved that malicious logic can be hidden in the development pipeline itself, and with LLMs, this danger intensifies through models whose behaviors emerge from opaque, hard-to-analyze weight matrices. Yet paradoxically, the very statistical nature of these models offers a path forward through ensemble consensus, creating a computational immune system where diverse models check each other's work. This shift from trusting individual black boxes to trusting their collective wisdom suggests a new paradigm for building robust AI systems - one that draws strength from the very complexity that makes them so challenging to analyze.

\bibliography{main}

\end{document}